# Capillary condensation under atomic-scale confinement

Qian Yang[1,2], P. Z. Sun[1,2], L. Fumagalli[2], Y. V. Stebunov[2], S. J. Haigh[3], Z. W. Zhou[4], I. V. Grigorieva[1,2], F. C. Wang[5,1], A. K. Geim[1,2]

[1]National Graphene Institute, University of Manchester, Manchester, UK
[2]Department of Physics and Astronomy, University of Manchester, Manchester, UK
[3]School of Materials, University of Manchester, Manchester, UK
[4]Key Laboratory of Advanced Technologies of Materials, School of Materials Science and Engineering, Southwest Jiaotong University, Chengdu, China
[5]Chinese Academy of Sciences Key Laboratory of Mechanical Behavior and Design of Materials, Department of Modern Mechanics, University of Science and Technology of China, Hefei, China

*Capillary condensation of water is ubiquitous in nature and technology. It routinely occurs in granular and porous media, can strongly alter such properties as adhesion, lubrication, friction and corrosion, and is important in many processes employed by microelectronics, pharmaceutical, food and other industries[1-4]. The century-old Kelvin equation[5] is commonly used to describe condensation phenomena and shown to hold well for liquid menisci with diameters as small as several nm[1-4,6-14]. For even smaller capillaries that are involved in condensation under ambient humidity and so of particular practical interest, the Kelvin equation is expected to break down because the required confinement becomes comparable to the size of water molecules[1-22]. Here we take advantage of van der Waals assembly of two-dimensional crystals to create atomic-scale capillaries and study condensation inside. Our smallest capillaries are less than 4 Å in height and can accommodate just a monolayer of water. Surprisingly, even at this scale, the macroscopic Kelvin equation using the characteristics of bulk water is found to describe accurately the condensation transition in strongly hydrophilic (mica) capillaries and remains qualitatively valid for weakly hydrophilic (graphite) ones. We show that this agreement is somewhat fortuitous and can be attributed to elastic deformation of capillary walls[23-25], which suppresses giant oscillatory behavior expected due to commensurability between atomic-scale confinement and water molecules[20,21]. Our work provides a much-needed basis for understanding of capillary effects at the smallest possible scale important in many realistic situations.*

The Kelvin equation predicts that capillaries become spontaneously filled with water at the relative humidity

$$RH_K = \exp(-2\sigma/k_B T d \rho_N) \qquad (1)$$

where $\sigma \approx 73$ mJ m$^{-2}$ is the surface tension of water at room temperature $T$, $\rho_N \approx 3.3 \times 10^{28}$ m$^{-3}$ is the number density of water, $k_B$ is the Boltzmann constant, and $d$ is the diameter of the meniscus curvature. For a two-dimensional (2D) confinement created by parallel walls separated by a distance $h$, $d = h/\cos(\theta)$ where $\theta$ is the contact angle of water on the walls' material. For capillary condensation to occur at relative humidity ($RH$) considerably below 100%, equation 1 dictates that $d$ must be comparable to $2\sigma/k_B T \rho_N \approx 1.1$ nm. For example, under typical ambient $RH$ of 40-50%, water is expected to condense in slits with $h < 1.5$ nm and cylindrical pores with diameters < 3 nm, if $\theta$ is close to zero. Even stronger confinement is required for capillaries involving less hydrophilic materials. So far, a broad consensus has been reached that the Kelvin equation remains accurate for menisci with $d \geq 8$ nm[1-4,6-11] and can also describe condensation phenomena in hydrophilic pores as small as 4 nm in diameter[12-14]. To achieve agreement with the experiments at this scale, the Kelvin equation is usually modified to account for so-called wetting films that are adsorbed on internal surfaces prior to the condensation transition and effectively narrow the capillaries. For the smallest capillaries, the thickness of the wetting films is used as a free parameter. In the real world, pores, cracks and cavities obviously do not terminate at the scale of several nm but extend even below 1 nm or $2\sigma/k_B T \rho_N$, the fact that makes condensation phenomena omnipresent under ambient conditions. The latter scale is comparable to the diameter of water molecules, which makes it challenging to study experimentally because of difficulties in creating the required atomic-scale confinement[1,10,12], the varying thickness of wetting films[1,2,7-13,17] and huge capillary pressures that can cause considerable deformations[13,23-25]. As for theory, the Kelvin equation is also believed to reach its applicability limit for confinement containing a few molecular layers because, at this smallest scale, the properties of water notably change[2,3,12,15,16] and the description in terms of homogeneous macroscopic thermodynamics becomes questionable[1-4,16-20], leaving aside the fact that such quantities as $d$ and $\theta$ in equation 1 can no longer be defined[1-3,18-20].

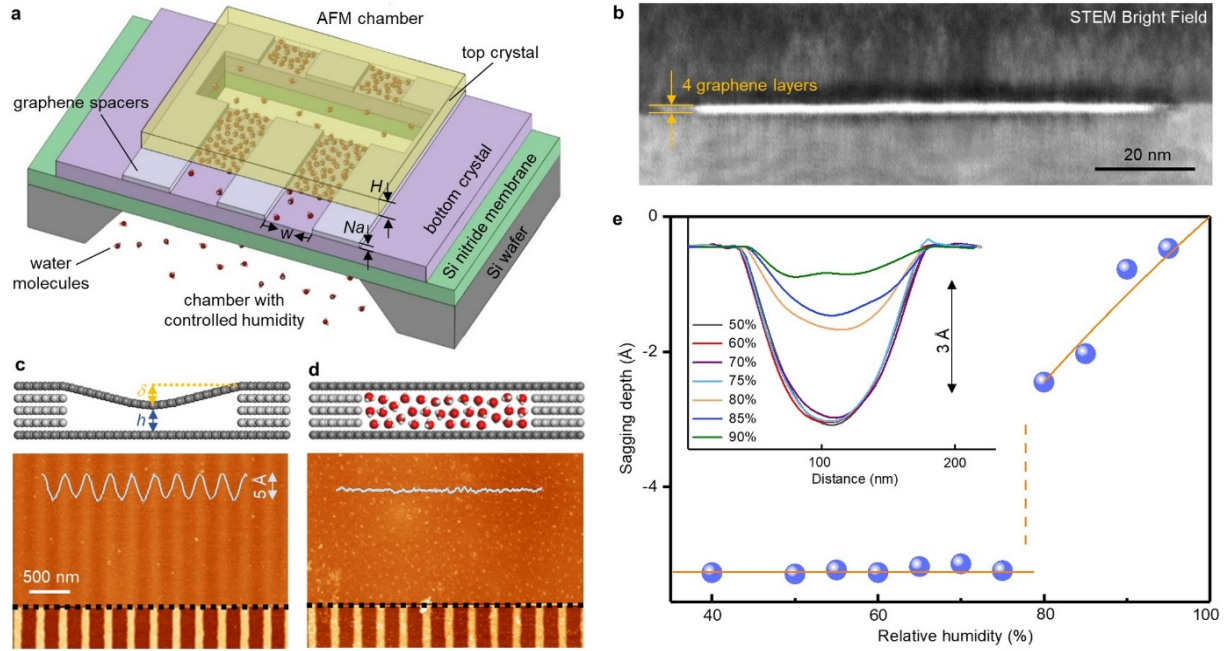

*Figure 1| Atomic-scale capillaries and water condensation inside. **a**, Schematic of the studied capillary devices. **b**, Cross-sectional imaging of a 4-layer graphite capillary, using scanning transmission electron microscopy. The top layer was >100 nm thick in this case. **c** and **d**, AFM imaging of the same mica capillary (N = 11) exposed to 30% and 95% relative humidity, respectively. In the dry state, the top crystal sagged by ~5 Å but became flat at high RH, as illustrated in the corresponding schematics above the images. The black dotted lines indicate the edge of the top crystal (cf. panel **a**). In the upper part of the AFM images, the color scale is given by the observed sagging (grey curves). The bottom part shows graphene spacers without the top crystal cover (dark-to-bright scale, 40 Å). **e**, Sagging depth $\delta$ as a function of RH for a graphite capillary with N = 4. Symbols: AFM measurements. The two solid curves in orange indicate the constant sagging $\delta_0$ below the condensation transition and the ln(RH) dependence above it. The transition is marked by the dashed vertical line. Inset: AFM profiles (averaged over 100 nm along the channel) of the top crystal for several values of RH. All the AFM measurements were carried out using the noncontact PeakForce mode ('AFM topography under controlled humidity' in SI).*

The studied capillary devices are shown schematically in Fig. 1a. Their most important part is atomically flat 2D channels made by van der Waals (vdW) assembly following the fabrication procedures described in Supplementary Information (SI). In brief, two atomically flat crystals were exfoliated from bulk muscovite mica or graphite to become top and bottom walls of our capillaries. Separately, narrow strips of multilayer graphene were fabricated to serve as spacers between the two mica or graphite crystals. Stacking the crystals and spacers on top of each other resulted in the 2D channels shown in Fig. 1a and *fig. S1*. We used graphene spacers between $N$ = 2 and >10 layers thick so that the capillaries had the designated height $Na$ (see Fig. 1a), where $a \approx 3.35$ Å is the effective thickness of monolayer graphene[26,27]. Examples of transmission electron microscopy imaging of our capillaries are provided in Fig. 1b and *fig. S1d*. Mica and graphite were chosen as archetypal strongly and weakly hydrophilic materials. Their contact angles are known to be in the range of 0–20° and 55–85°, respectively[16,28,29]. For surfaces exposed to air, $\theta$ is close to the above upper bounds[28,29] (Supplementary Information).

To detect *RH* at which capillary condensation occurred in the described 2D capillaries, we exploited the fact[26,30] that suspended thin crystals exhibited noticeable sagging caused by their vdW adhesion to sidewalls (Fig. 1c). In our experiments we found that, when the capillaries became filled with water, the sagging depth $\delta$ diminished (Fig. 1d), presumably because intercalating water molecules 'screen' the adhesion[27,30]. To make the resulting changes in $\delta$ detectable by atomic force microscopy (AFM), it was important to carefully choose the top crystal's thickness *H* and the channel width *w* (see Fig. 1a). As described in Supplementary Information ('Remnant sagging above the condensation transition'), these two parameters define the stiffness of the top crystal and, hence, how deep it bends inwards. We found that, for $w \approx 150$ nm, the top crystal should be ~50–70 nm thick to exhibit the sagging depth $\delta$ of several Å. If either *w* or *H* were changed only by a factor of 2, the strong dependence $\delta \propto w^4/H^3$ resulted in either collapsed channels (the top crystal attached to the bottom one) or such small $\delta < 1$ Å that the condensation transition was impossible to discern by AFM. The studied capillaries were typically 5 to 10 μm long.

As shown in Fig. 1a and *fig. S1c*, our capillary devices were assembled on top of a silicon nitride membrane. It had a rectangular opening that was extended into the bottom crystal by dry etching. The Si chip supporting the entire assembly was used to separate two miniature gas chambers that were integrated into an AFM setup as shown in *fig. S2a*. The bottom chamber provided variable humidity so that one entrance of 2D capillaries was exposed to chosen *RH*. The opposite entrance was facing the top chamber, which enclosed the AFM scanning head and was usually kept at low humidity. The two-chamber configuration allowed us to avoid the influence of *RH* on the measurements of the top crystal's topography (e.g., no condensation occurred at the AFM tip during scanning)[31]. Examples of AFM imaging for mica and graphite devices are given in Figs. 1c-d and *fig. S3*, respectively. They reveal pronounced sagging under dry conditions, which disappeared in high humidity. Typical evolution of the top crystal's profiles with changing *RH* is shown in Fig. 1e and *figs. S4,S5*. In these measurements, we increased *RH* inside the bottom chamber in steps of 5%, waited for an hour for the system to stabilize and then recorded AFM images. The temperature was kept at 294 ±1 K. For the device in Fig. 1e, the sagging remained practically constant for *RH* ≤75 % and then exhibited a pronounced jump at $RH_C$, which we attribute to the condensation transition (similar behavior is shown in *fig. S5*). Further increase in *RH* led to a gradual decrease in $\delta$ such that the top crystal became practically flat at *RH* > 95% (Fig. 1). The remnant sagging at *RH* > $RH_C$ is well described by the negative capillary pressure that keeps the top crystal bent inwards even after water filled the 2D channels suppressing the adhesion of the top crystal to the sidewalls. Indeed, for *RH* > $RH_C$, $\delta$ is expected to evolve as $\propto \ln(RH)$ and reach zero at 100% humidity[23,24], in agreement with the observed behavior in Fig. 1e and *fig. S6* ('Remnant sagging above the condensation transition'). If we repeated the same measurements but with decreasing *RH*, a reverse jump occurred at the same $RH_C$, that is, the condensation transition was non-hysteretic (*fig. S4a*; 'Non-hysteretic behavior of the condensation transition'). Note, however, that it could take up to several days for capillaries exposed to high *RH* to completely dry out and return to their original state (*fig. S4b*). On the other hand, for measurements with increasing *RH*, no difference in $RH_C$ was observed after either an hour or days of equilibration. Accordingly, our experiments were normally carried out with increasing rather than decreasing *RH*, as in Fig. 1e.

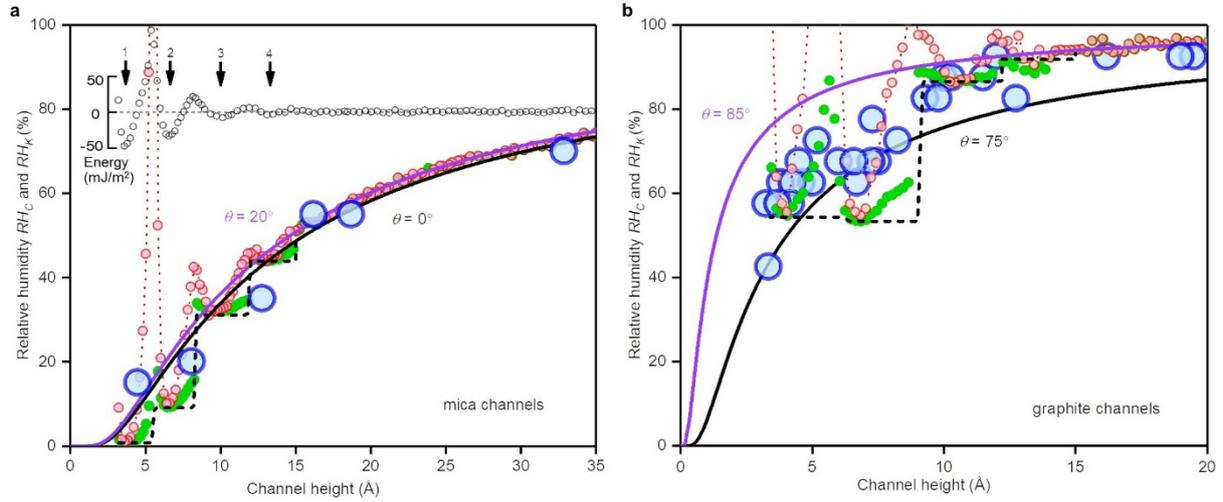

***Figure 2| Condensation transition under extreme 2D confinement. a,*** *Relative humidity $RH_C$ required for water condensation in mica channels of different heights h. Experimental observations: blue circles; their size reflects the 3.5% experimental uncertainty in determining $RH_C$ ('AFM topography under controlled humidity' in SI). Two solid curves: $RH_K$ given by equation 1 with bulk water's characteristics for the range of possible $\theta$ for mica (color coded). The upper curve (black circles) with its own Y-axis and the common X-axis shows our MD calculations for changes in $\gamma_{SL}$ caused by restructuring of water inside 2D channels ($\theta \approx 10°$). The arrows mark the energy minima that correspond to the integer number of water monolayers that can fit inside the 2D capillaries. Red dotted curve with symbols: the effect of the oscillating $\gamma_{SL}$ on the condensation transition, according to equation 2. Black dashed curve: same analysis but assuming fully flexible capillary walls allowing relaxation into the energy minima at commensurate h. Green solid circles: same analysis but for a finite rigidity of the confining walls.* ***b,*** *Same as in **a** for graphite capillaries. The simulated curves are for $\theta \approx 85°$.*

Figure 2 summarizes our results for the condensation transitions observed in mica and graphite 2D capillaries. To allow more accurate comparison between data collected from different devices, we have accounted for the fact that capillaries with the same *N* often exhibited different sagging in their dry state, $\delta_0$. For capillaries with large $\delta_0$, we observed consistently lower $RH_C$ than for those with small initial sagging and same *N*. Moreover, comparing capillaries with different *N* but similar channel heights $h = Na - \delta_0$, we found close values of $RH_C$ (*fig. S5*). This implies that it was the narrowest, central region of the 2D channels which determined the onset of condensation, in agreement with general expectations[32] ('Effect of initial sagging'). Accordingly, to account for the effect of different $\delta_0$, Fig. 2 plots $RH_C$ as a function of *h* rather than *N*. For mica capillaries, the experimental data are well described by equation 1 using $\theta$ and $\sigma$ of bulk water. Because $RH_K(h)$ depends little on the exact value of $\theta$ for strongly hydrophilic capillaries (Fig. 2a), the comparison of $RH_C$ for mica with equation 1 is straightforward. This is not the case for weakly hydrophilic graphite where relatively small variations in $\theta$ lead to considerable changes in $RH_K(h)$ as per equation 1. Nonetheless, the values of $RH_C$ observed for our graphite capillaries fall well within the range expected from the Kelvin equation using the contact angles $\theta = 80 \pm 5°$, typical for graphite surfaces under ambient conditions[29].

It is surprising that the macroscopic Kevin equation using the characteristics of bulk water describes so well condensation in our mica capillaries and, also, provides qualitative agreement for the graphite capillaries. As mentioned in the introduction, strong discrepancy is expected for the angstrom-scale confinement where only 1 or 2 layers of water fit inside capillaries. Before trying to explain the unexpected agreement between the experiment and the macroscopic Kelvin equation, we note that $RH_C$ in Fig. 2a are notably lower than the *RH* values required to achieve condensation in the previous studies for $d \geq 8$ nm. At our low *RH*, no continuous wetting layer is expected even on fresh mica surfaces[12,33], and a partial coverage by monolayer water is probably suppressed further by adsorbates from air, which are responsible for the relatively large $\theta$ close to 20°. The same consideration about the apparent absence of wetting films also applies for the graphite capillaries in which the wetting transition is even less likely[1,2,28]. Second, to avoid the macroscopic variables $\sigma$ and $\theta$ that are poorly defined under our extreme confinement, the Kelvin equation can be rewritten as[1,2,18]

$$RH_K = \exp[-2(\gamma_{SV} - \gamma_{SL})/hk_BT\rho_N] \qquad (2)$$

where $\gamma_{SV}$ and $\gamma_{SL}$ are the surface energies for solid-vapor and solid-liquid interfaces, respectively, and $\gamma_{SV} - \gamma_{SL} = \sigma\cos\theta$. The former energy $\gamma_{SV}$ is largely independent of $h$ because the interaction of gas molecules with surfaces should depend little on confinement. Also, $\rho_N$ changes relatively little for nearly incompressible water[20,34]. Therefore, the dominant effect of extreme confinement is likely to come from $\gamma_{SL}(h)$ that is governed by vdW interactions of liquid water with solid surfaces. Because these interactions are short-range, it is predominantly the first near-surface layer of water that determines $\gamma_{SL}$. If this layer changes little under confinement, then $\Delta\gamma = \gamma_{SL}(h) - \gamma_{SL}(\infty) \approx 0$ and capillary condensation should closely follow equation 1 even at nanoscale[1,2,18]. Substantial changes in $\gamma_{SL}$ and, hence, $RH_C$ are expected only in the limit of few-layer water where its near-surface structure is notably altered[20,34] (fig. S7).

For further analysis, we employed molecular dynamics simulations (SI) to evaluate $\Delta\gamma$ and the resulting corrections to the macroscopic Kelvin equation, which are given by the factor $\exp(2\Delta\gamma/hk_BT\rho_N)$ according to equation 2. Examples of the calculated $\Delta\gamma(h)$ are shown in Fig. 2a and *fig. S8*. There are pronounced commensurability oscillations[20,21,34] in $\gamma_{SL}(h)$ such that energy minima appear if 2D channels accommodate exactly 1, 2, 3 or 4 layers of water. The oscillations practically disappear for $h > 15$ Å where $\Delta\gamma$ becomes almost zero, which also implies that the macroscopic Kelvin equation should be valid in this regime. For smaller $h$, changes in $\gamma_{SL}$ are comparable to $\sigma$, which means that the above correction factor is comparable to $RH_K$ itself. Consequently, the simulated $RH_C(h)$ dependences shown in Fig. 2 (red curves) exhibit giant oscillations such that, for incommensurate $h$, water condensation becomes unfavorable and should not occur even at 100% humidity. No such oscillatory behavior could be detected in our experiments. We attribute its absence to elastic adjustment such that 2D channels tend to accommodate an integer number of molecular layers of water[20]. Indeed, the energy minimization should be applied to the entire system, including the elastic energy of confining walls[23-25]. For an extremely soft confinement, 2D channels would adjust their $h$ to reach the commensurate states at minima of $\Delta\gamma$. The condensation behavior in this case should follow the step-like black curves shown in Fig. 2. A finite rigidity pushes the equilibrium conditions away from the commensurability minima. To estimate a likely elastic response of our 2D channels, note that the capillary pressure above $RH_C$, which is defined by $\sigma$, keeps the top crystal bent inwards typically by several Å (Fig. 1e; *figs. S4-6*). Similar elastic adjustments can be expected in our capillaries because changes in $\Delta\gamma$ are comparable to the absolute value of $\sigma$. Accordingly, our confinement should be considered as rather soft. To illustrate the likely condensation behavior in such a case, the green curves in Fig. 2 show the $RH_C(h)$ dependences expected if the walls' finite rigidity allows their deformations to reach within 0.5 Å from the commensurability minima in $\Delta\gamma$. The latter curves are in good agreement with the experiment and, in the case of graphite capillaries, also exhibit the same trend towards lower $RH$ for $h < 10$ Å as observed experimentally in Fig. 2b. In principle, the elastic response of 2D confinement could be included in the simulations self-consistently but the scatter in the experimental data and different $H$ used for different capillary devices make this effort beyond the rationale of the present study.

Finally, we note that elastic adjustments should also play an important role in real-life capillaries responsible for condensation phenomena under ambient humidity. Indeed, capillary pressures at 1-nm scale typically exceed 1 kbar and the resulting elastic response of even infinitely thick walls can exceed 1 Å for the case of 2D confinement ('Remnant sagging above the condensation transition'). This should force atomic-scale capillaries to elastically adjust their geometry[13,23,24], suppressing commensurability oscillations and resulting in the condensation transition at $RH$ close to the values prescribed for a soft confinement. Accordingly, capillary condensation under ambient conditions can be expected to qualitatively follow the macroscopic Kelvin equation, as happened for the reported capillaries.

**Supplementary Information**

**Fabrication procedures.** The studied capillary devices were fabricated following the procedures described in refs. 26&27 and shown in the flow chart of *fig. S1a.* First, a large crystal of multilayer graphene (the number of layers, *N*) was prepared on an oxidized Si wafer by mechanical exfoliation. Using e-beam lithography and oxygen plasma etching, the crystal was patterned into a set of parallel narrow strips that had a width of ~150 nm and were separated by approximately the same distance (*fig. S1b*). These spacers were then put on top of a mica or graphite crystal of typically 20 nm in thickness. The latter crystal was prepared on a separate Si wafer and, in this work, is referred to as bottom crystal (*fig. S1a*①).

In parallel, we prepared a suspended silicon nitride (SiN) membrane with a rectangular hole in the center (*fig. S1a*②). To this end, we used commercial silicon wafers with 500 nm of SiN deposited on both sides. Using photolithography and reactive ion etching (RIE), a window of about 750×750 $\mu m^2$ in size was made in one of the SiN layers. The wafer was then placed in hot KOH to etch through the entire Si thickness and obtain a freestanding SiN membrane of ~70×70 $\mu m^2$ in size. After that, a rectangular hole (~3×20 $\mu m^2$) was plasma-etched in the SiN membrane, using another round of photolithography and RIE (*fig. S1a*②).

The two-layer assembly consisting of the bottom crystal and graphene spacers (*fig. S1a*①) was transferred on top of the SiN membrane in such a way that graphene strips were aligned perpendicular to the long edge of the rectangular hole. This step was followed by RIE from the backside of the Si wafer to extend the hole through the bottom crystal (*fig. S1a*③). Finally, another mica (or graphite) crystal was placed on top of the two-layer assembly to form 2D channels (*fig. S1a*④, *1c*).

After each crystal transfer, samples were cleaned in acetone, deionized water and isopropanol. This was followed by annealing at 400°C in a hydrogen-argon atmosphere for 3 h. Such thorough cleaning was essential to remove polymer residues and other possible contamination, which could otherwise block the capillaries. In our experiments, we used only those capillaries that exhibited uniform sagging along their entire length, such as those shown, for example, in Fig. 1c of the main text and *fig. S3.*

The contact angle for muscovite mica and natural graphite used for making our devices was measured by Drop Shape Analyzer 100S from KRÜSS. We found $\theta \approx$ 80-85° for graphite and 15-20° for mica after exposure to ambient air for a few days, in agreement with the previous reports (see, e.g., refs. 28&29).

**AFM topography under controlled humidity.** Our setup for AFM measurements is shown in *fig. S2a*. The SiN wafer containing a capillary device such as shown in *fig. S1c* was placed to seal an airtight metal chamber with a volume of about 1 $cm^3$. A continuous flow of nitrogen gas was provided into this chamber through miniature inlets. The humidity was controlled by mixing dry nitrogen with nitrogen bubbled through deionized water, using a computer-controlled flowmeter. *RH* was measured by a humidity sensor (Sensirion), which was mounted inside the bottom chamber. The sensor was calibrated using 3 different saturated salt solutions (lithium chloride, magnesium nitride and potassium chloride) to ensure readings of *RH* with an accuracy of ±1%. A commercial silicone rubber enclosure (from Bruker) was attached to the AFM head (*fig. S2a*). When it was lowered for taking AFM images, the enclosure edges sealed the space above the studied devices. Fresh silica gel granules were usually placed inside the enclosure to provide low humidity in the top chamber.

All AFM images in our experiments were taken in the PeakForce mode using Dimension FastScan from Bruker and analysed using WSxM software[35]. Selected capillaries were imaged at regular *RH* intervals of 5%, after stabilizing humidity in the bottom chamber for approximately an hour. As an example of our AFM measurements, *fig. S3* shows three critical steps in the evolution of the topography for an *N* = 3 graphite capillary. The initial sagging in this case was ~4 Å as seen for *RH* = 55% in *fig. S3a*. With increasing *RH* in 5% steps, no change in the sagging profile was observed until we reached 70% *RH*. At the latter humidity, the top crystal was found to be notably less sagged (*fig. S3b*). With increasing *RH* further, the top crystal gradually lifted and became practically flat at 95% *RH* (*fig. S3c*). The flattening process is described in detail below. Because the rapid change in sagging happened somewhere between 65 and 70% *RH*, we assigned the condensation transition to *RH$_C$* = 67.5 ±2.5% with an additional error of ±1% because of the humidity sensor's accuracy (error bars in Fig. 2 of the main text).

The use of low humidity in the top chamber notably improved the stability of AFM imaging but was not essential. Indeed, if we simply connected the top and bottom chambers so that both sides of the studied capillaries were exposed to the same *RH*, the condensation transition was found to occur at the same *RH$_C$* as in the case of low *RH* in the top chamber. This shows that the condensation transition is determined by the *RH* at the entrance side (that is, the highest *RH*) (*fig. S2b*). This observation is consistent with the fact that no difference in sagging was observed along the entire length of the capillaries, even when their exits were at low *RH* (Fig. 1d of the main text; *fig. S3b*), which indicates no detectable gradient in negative capillary pressure along the 2D channels. The constant capillary pressure can be attributed to a very fast flow of liquid water through our atomically flat capillaries[26], which allows essentially the same meniscus curvature at both entrance and exit sides, as shown schematically in *fig. S2b*. If the top chamber is kept at low humidity, such an equilibrium state is stabilized by a slightly retracted exit meniscus and slow Knudsen diffusion of water vapor near the capillary exit[26], which provide the required *RH* gradient. This is different from the case of nanoporous media with rough internal surfaces and tortuous capillaries, where both liquid and vapor transport are slow, allowing large pressure gradients to build up along the liquid flow direction[14].

**Non-hysteretic behavior of the condensation transition.** Capillary condensation in nanochannels is often accompanied by hysteresis such that *RH$_C$* required to reach the transition depends on whether external *RH* is increased or decreased[1,4,11,13,23,24,36]. This was not the case of our capillaries, which exhibited no hysteresis within our experimental accuracy. This behavior is illustrated in *fig. S4a*, which shows top crystal's profiles for a 4-layer graphite capillary where *RH* was changed in a small loop around the transition observed at *RH$_C$* = 77.5 ±2.5%. The capillary's sagging was constant for *RH* ≤ 75%. Then we increased *RH* to 80% and equilibrated for 1 h, following the experimental procedures described above. The 5% change in *RH* led to a pronounced jump in the sagging depth $\delta$, indicating the condensation transition (cf. black and red curves in *fig. S4a*). The capillary profile remained stable while *RH* was maintained at 80%. When we decreased *RH* back to 75%, the capillary did not return to the initial dry state after 4 h (blue curve). Nonetheless, the top crystal continued to sag gradually with time (*fig. S4a*). The dry state was eventually reached (after more than 9 but less than 16 h). Therefore, the condensation transition occurred at the same *RH$_C$* with either increasing or decreasing humidity, although long equilibration times were needed for 2D channels to dry up.

The slow recovery of the initial dry state is further exemplified by *fig. S4b*. It shows the case of a graphite capillary with *N* = 6, where the condensation transition was found to occur between 60 and 65% *RH*. In *fig. S4b*, we first increased *RH* from 60% directly to 95%, well above the transition (black and red curves, respectively). Then *RH* was reduced to ~30%, well below *RH$_C$* = 62.5 ±2.5%. As seen in *fig. S4b*, the top crystal regained its original profile very slowly, and the capillary returned to its dry state only after several days. After this the sagging remained stable. The reason for such a slow drying process remains to be understood.

It is also worth mentioning that our capillary devices did not show any discernable change in sagging with changing *RH* below the condensation transition, as seen, e.g., in Fig. 1e. This is in contrast to the usual elasto-capillary response of nanoporous media, in which adsorption of water molecules on internal surfaces leads to notable strain, usually referred to as the Bangham effect[37]. Its apparent absence in our experiment is perhaps not surprising. First, as discussed in the main text, we expect only a small partial coverage of internal walls by adsorbed water molecules before the transition. Second, if there were significant adsorption, the adsorption-induced strain is typically of the order of $10^{-4}$ for materials with high Young's moduli[13,23,24,36]. Therefore, for our top crystals with *H* and *w* of ~100 nm, this strain would translate into sagging of the order of 0.1 Å, below our experimental accuracy.

**Effect of initial sagging.** For 2D channels with the same $N$, their heights $h = Na - \delta_0$ could vary considerably because of different $H$ and slightly different $w$, which control the initial sagging $\delta_0$. This resulted in different $RH_C$ observed for capillaries with the same $N$. This behavior is illustrated in *fig. S5* showing the condensation transition in two capillaries with $N = 5$ but different $\delta_0$. The capillary in *fig. S5a* had the top layer with $H \approx 70$ nm and exhibited initial sagging of ~4 Å. The transition in this device occurred at $RH_C$ = 82.5% as seen in *fig. S5a*. The other capillary (*fig. S5b*) had a thinner (~45 nm) top crystal and, accordingly, its $\delta_0$ was larger (~ 8.5 Å). The latter device exhibited $RH_C$ = 72.5%, considerably lower than that in *fig. S5a*. This implies that $N$ was not the characteristic determining the onset of water condensation. The importance of $h$ rather than $N$ is even better exemplified by the results of *fig. S4*. Indeed, the 4-layer capillary here exhibited the condensation transition at 77.5% RH whereas the nominally larger capillary with $N = 6$ in *fig. S4b* showed the transition at lower $RH_C$ = 62.5%. This obviously contradicts the general expectations that the smaller $N$ capillary should exhibit lower $RH_C$. However, because of different $\delta_0$, the smaller ($N = 4$) capillary in *fig. S4* had $h \approx 7$ Å, which was larger than $h \approx 3.5$ Å for the larger ($N = 6$) capillary. The smaller $RH_C$ for 2D channels with smaller $h$ agrees with the general expectations.

The above observations strongly suggest that $h$ is the size parameter that describes best the condensation transition in our atomic-scale capillaries. This means that it is the central region between the sagged-top and flat-bottom crystals where the condensation process is effectively initiated. This is not entirely unexpected because molecular dynamics (MD) simulations previously showed that corner menisci in narrow channels were unfavorable for condensation[32]. Furthermore, note that the mean free path of water molecules in air is about 65 nm, which is comparable to the channel width $w \approx 150$ nm. This implies that the entire channel should present a single entity from the standpoint of thermodynamics, allowing only one condensation transition over the channel's cross-section. To this end, it is important to note that, although our capillaries contained only a few monolayers of water, there were still millions of molecules inside each capillary, which should be sufficient for the thermodynamic description, unlike in the case of nm-scale droplets containing a small number of molecules[3,16,18].

**Remnant sagging above the condensation transition.** Nanometer-thick suspended crystals are known to exhibit considerable sagging, which is believed to be caused by their vdW adhesion to sidewalls[26,30,38,39]. As water spontaneously condensed inside the 2D channels, the sagging was found to suddenly decrease (in a jump-like fashion as seen in Fig. 1e of the main text and *figs. S4a* and *S5*), which we attribute to suppression of the vdW adhesion by intercalating water molecules. This explanation is supported by MD simulations[27]. They showed that capillaries with monolayer spacers ($N = 1$) always collapsed (independently of $H$) because of vdW interaction between the top and bottom crystals. The collapsed capillaries could be opened by intercalating water, because the attraction rapidly diminishes at distances more than a few Å so that even a monolayer of water provided sufficient 'screening' of the vdW attraction[26,27].

In all our measurements, the sagging depth $\delta_0$ became abruptly smaller at the condensation transition but did not completely disappear. Only if $RH$ was increased further, the remnant sagging gradually decreased, approaching zero at 100% $RH$, so that the top layer became essentially flat (Figs. 1d,e of the main text; *fig. S3*). The remnant sagging at the transition and its gradual changes with increasing $RH$ further can be explained by the negative pressure $P$ caused by the condensed water meniscus[1,40]. Indeed, let us consider our typical mica capillary having $w \approx 150$ nm and the top layer's thickness $H \approx 50$-60 nm (*fig. S6a*). After the condensation transition occurred at a certain $RH$ (that depended on the channel height $h$), the top layer remained sagged typically by several Å. At the condensation transition the capillary pressure is given by $P \approx 2\sigma \cos\theta/h$. Using the contact angle $\theta \approx 20°$ for mica and the surface tension of bulk water, $\sigma \approx 73$ mN/m, the Young-Laplace equation yields $P \approx 700$ bar for $h \approx 2$ nm.

The negative pressure $P$ forces the top crystal to bend downwards, resulting in its sagging given by[41]

$$\delta = \frac{5Pw^4}{32EH^3} \quad (3)$$

where $E \approx 60$ GPa is Young's modulus of mica in the out-of-plane direction[42]. Equation 3 yields $\delta \approx 4 - 7$ Å, in agreement with our observations. This substantiates our model that the remnant sagging above the condensation transition is caused by the negative capillary pressure. Similar agreement is found for graphite capillaries, although there is a larger uncertainty in the estimates (a factor of 2) because $P$ strongly depends on $\theta$ for the large contact angles exhibited by water on graphite. As $RH$ is increased beyond the condensation point, the meniscus extends outside capillaries, and its curvature becomes progressively smaller to match the external $RH$. Accordingly, the negative capillary pressure above $RH_C$ evolves with $RH$ and is given by the Kelvin equation as[1,23,24,36,40]

$$P = k_B T \rho_N \ln(RH) \quad (4).$$

According to this equation, the pressure that bends the top crystal should decrease logarithmically with $RH$, in good agreement with our observations (*fig. S6b*). Note that, close to 100% $RH$, $\delta \propto P \propto \ln(RH) \approx (RH - 1)$ is expected to approach zero linearly, as indeed observed in Fig. 1e and *fig. S6b*.

The condition of partially sagged but open capillaries (that is, a few Å $< \delta_0 < Na$, as in our devices) is rather difficult to satisfy experimentally. Indeed, if we were to decrease $H$ or increase $w$ by only a factor of 2 with respect to the found optimal design, $\delta$ in equation 3 would increase by an order of magnitude because of the high powers. On the other hand, the capillary pressure $P$ in equation 4 depends on $RH$ only logarithmically, which means that even at very low humidity (e.g., 5%), it would be thermodynamically favorable for the top layer with the non-optimal $w$ or $H$ to bend all the way down and reach the channel's bottom. Therefore, such non-optimized 2D channels are unstable with respect to spontaneous water condensation under low-humidity conditions. If we were to do the opposite and increase the top crystal's stiffness (by using twice smaller $w$ or twice larger $H$), $\delta$ in equation 3 becomes so small ($< 1$ Å) that changes in the sagging would be impossible to detect by AFM. The above consideration shows that there is a subtle interplay between materials parameters and 2D channels' design, and stringent rules should be followed in order to detect the condensation transition in experiment. Following this insight, we usually increased $H$ by ~50% for our smallest 2D channels with $N = 2$ and 3, which ensured that they remained open. Also, when making graphite capillaries, we used top crystals slightly (~20%) thicker than in the case of mica capillaries with the same $N$ because mica has a higher Young's modulus than graphite[43].

For nanoscale 2D capillaries such as cracks or slits inside bulk materials ($H \to \infty$), their elastic deformations caused by large capillary pressures can notably shift the condensation transition with respect to that expected for the rigid confinement[20,21]. To estimate the magnitude of such adjustments, let us consider the deformation of a half-space elastic medium subject to the uniform load $p$ over a suspended strip with the width $w = 2a$ in the range of $-a \leq x \leq a$. The vertical deformation is given by[44]

$$u_z(x) = -\frac{(1-\nu)p}{\pi G} a \left[ \left(\frac{x}{a} + 1\right) \ln|x + a| - \left(\frac{x}{a} - 1\right) \ln|x - a| - 2 \right] \quad (5)$$

where $\nu$ is Poisson's ratio and $G$ is the shear modulus. This equation yields the sagging

$$\delta = u_z(0) - u_z(\pm a) = \frac{\ln(2)(1-\nu)pw}{\pi G} \quad (6).$$

If we take as an example the elastic properties of graphite with $G \approx 10$ GPa and $\nu \approx 0.3$[43], equation 6 yields $\delta \approx 2.3$ Å for capillary pressures of about 1,000 bar. Such $P$ are typical for cavities of 1-2 nm in height (see above). This indicates that elastic deformations can be not only a contributing factor during the condensation transition[23,24,36] but also allow atomic-scale cavities in bulk materials to adjust their size so that an integer number of water layers can fit inside, similar to our case where the top crystal was intentionally made sufficiently flexible.

**Molecular dynamics simulations of water-surface interaction under strong confinement.** To investigate the dependence of the solid-liquid surface energy $\gamma_{SL}$ on $h$, MD simulations were performed using LAMMPS[45] and the SPC/E model for water molecules[46]. The interaction between water and confining walls was modeled using the Lennard-Jones potential with parameters taken from Ref. 47. Flat rigid graphene sheets were used to mimic the confining walls. For simplicity, to account for surfaces with different $\theta$, we varied the interaction energies of carbon with hydrogen $\varepsilon_{HC}$ and oxygen $\varepsilon_{OC}$. These energies[47] were multiplied by a factor of $k$ that was varied from 0.7 to 1.3 in steps of 0.2 to find the water-wall interaction that would approximate the experimental contact angles. The MD angles $\theta$ were estimated using water droplets containing 4,000 molecules. Our simulations yielded $\theta \approx 85°, 63°, 30°$ and $11°$ for $k = 0.7, 0.9, 1.1$ and $1.3$, respectively. The insets of *fig. S7* show the profiles for the water droplets found in the case of $\theta \approx 11°$ and $85°$. We used these two $\theta$ and the corresponding $k$ to model $\gamma_{SL}(h)$ for our mica and graphite capillaries, respectively. Note that the former value lies in the middle of the contact-angle interval observed for mica[28] and, importantly, our MD results exhibited little sensitivity to the exact $\theta$ for strongly hydrophilic capillaries, as expected from the $\cos(\theta)$ dependence.

Having established parameters for the desired contact angles, we proceeded to another simulation setup that consisted of two flat 4-layer graphite sheets immersed in a water box containing 40,000 molecules. The dimension of each graphene sheet was $102.2 \times 100.9$ Å$^2$ whereas the water box was $140.0 \times 140.0$ Å$^2$ in size, which allowed water molecules confined between the rigid graphite sheets to exchange easily with outside molecules. After an equilibration run of 1.0 ns, the two sheets were brought progressively closer in steps of 0.2 Å. Each time the system was equilibrated for 0.1 ns and its total potential energy was calculated for further analysis. Periodic boundary conditions were imposed in all three directions. All the simulations were carried out with the isothermal-isobaric ensemble at 298 K. The density profiles found in our simulations are shown in *fig. S7*. The confined water exhibits a pronounced layered structure that extends over 2 intermolecular distances from each surface, before the water density converges to its bulk value, in agreement with the earlier literature (see, e.g., refs. 19, 20, 48, 49).

The deviations $\Delta\gamma$ in the solid-liquid surface energy $\gamma_{SL}$ from its bulk value may be considered as extra work spent to rearrange water molecules into the strongly layered structures shown in *fig. S7*. If $h$ is sufficiently large, the extra work is negligible because the opposite surfaces do not 'feel' each other, and their near-surface water structures remain unchanged with respect to the case of infinite $h$. However, as the walls are getting closer, the layered structures overlap (see the density profiles for $h$ < 10 Å in *fig. S7*). As a result, the total energy and, hence, $\Delta\gamma$ exhibit pronounced oscillations (*fig. S8*). Using equation 2 and the numerically found $\Delta\gamma$, it is straightforward to calculate the *RH* required for water condensation inside atomic-scale capillaries. The results are plotted in Fig. 2 of the main text and reveal giant oscillations in $RH_C$ which emerge when the structured layers of water near the two confining surfaces start overlapping. Note that the confining walls in the MD simulations were made rigid disallowing elastic deformations considered separately in our analysis in Fig. 2.

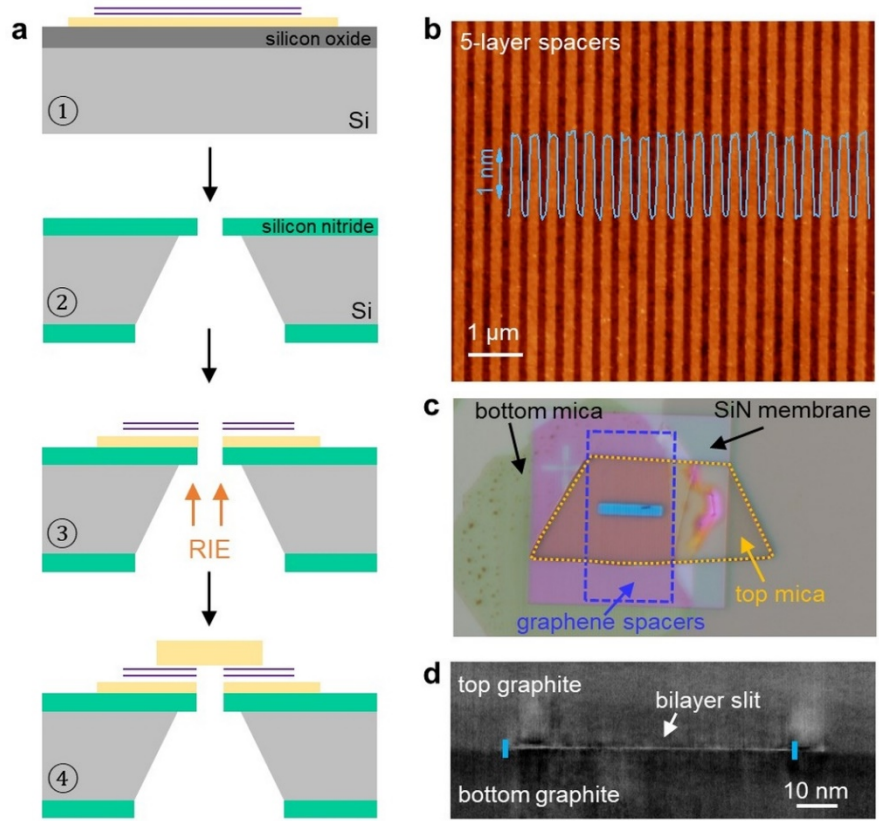

**Figure S1 | Nanofabrication of two-dimensional channels. a,** Simplified flow chart for our fabrication procedures. (1) Graphene spacers and the bottom crystal of either mica or graphite (shown in yellow) were assembled on top of an oxidized Si wafer. (2) A suspended SiN membrane with a rectangular hole was prepared separately. (3) The two-layer assembly was transferred from the Si oxide wafer onto the SiN membrane. The opening was extended through the assembly by RIE. (4) The top crystal of mica or graphite was placed on top of graphene spacers. **b,** AFM micrograph of graphene spacers with N = 5. The color scale is given by the height profile (blue curve). **c,** Optical image of a final mica device used in our experiments. The bottom mica crystal shows up in purple on top of the square SiN membrane. Graphene spacers (N = 3) and the top mica layer are contoured in blue and yellow, respectively. **d,** Cross-sectional STEM image of a graphite channel with N = 2. The blue ticks mark the channel's edges.

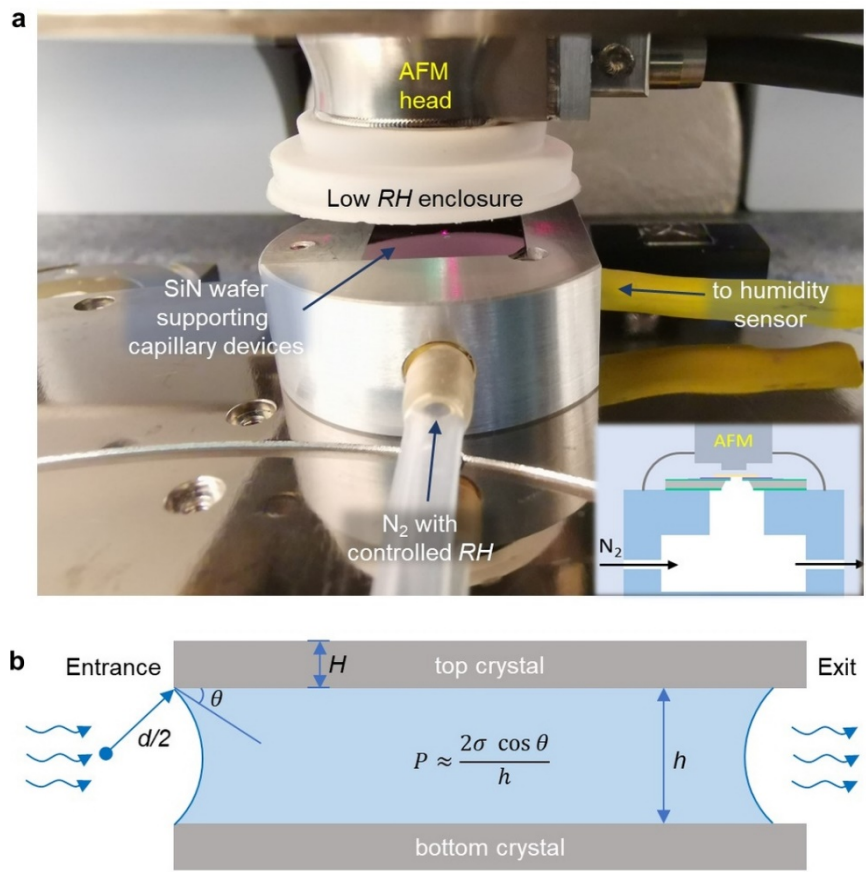

**Figure S2 | Measurements of capillary condensation. a**, Our AFM setup. Humidified nitrogen gas flows through the bottom chamber made from an aluminum alloy. A silicon wafer of 15×15 mm² in size is seen to cover the chamber, flush with its top surface. The white rubber gasket was lowered down during AFM measurements to seal the space above the Si wafer. Inset: cross-sectional schematic showing how capillary devices were mounted during AFM measurements. **b,** Schematics of a water plug inside our capillaries. For brevity, the layered structure of water is ignored in this sketch. When the top chamber is at low RH, the meniscus slightly retracts inside the capillary to create a vapor pressure gradient. The RH gradient stabilizes two menisci with the same curvature at both exit and entrance. The distance from the exit meniscus to the opening is expected to be short because, in our atomically flat capillaries, water moves much faster as liquid than vapor[26].

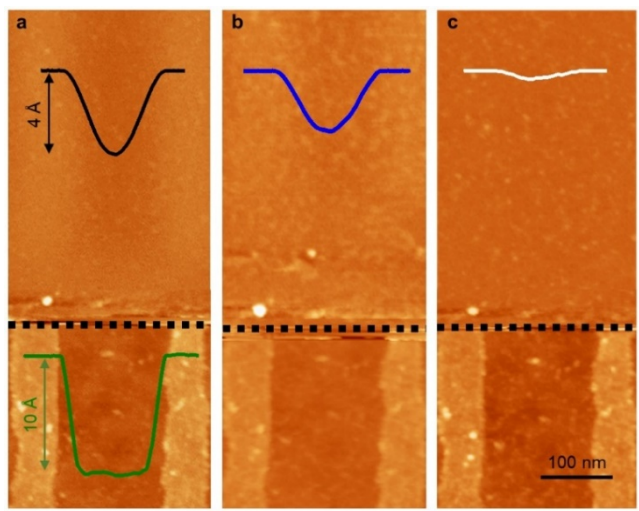

**Figure S3 | Visualization of the condensation transition using AFM.** Images of a graphite capillary with N = 3 at RH of 55, 70 and 95% (**a**, **b** and **c**, respectively). The upper parts of the images show sagging of the top graphite crystal (H ≈ 80 nm) into the 2D channel. The lower parts are the area immediately outside the channel, which is not covered by the top graphite. The black dotted lines mark a border between the two regions (edge of the top crystal). The color scales for the lower and upper parts of the AFM images are given by the green and black curves, respectively. The profiles are averaged over ~100 nm along the Y-direction, and the curves in the upper parts of all the panels are provided in the same scale given by the black arrows in panel **a**. A small number of horizontal scanning lines (X-direction) around the black-dot dividing lines were removed for clarity because they contained numerous instabilities caused by the AFM tip moving along the edge of the top crystal and jumping up and down. Such instabilities are typical for AFM scanning close to edges.

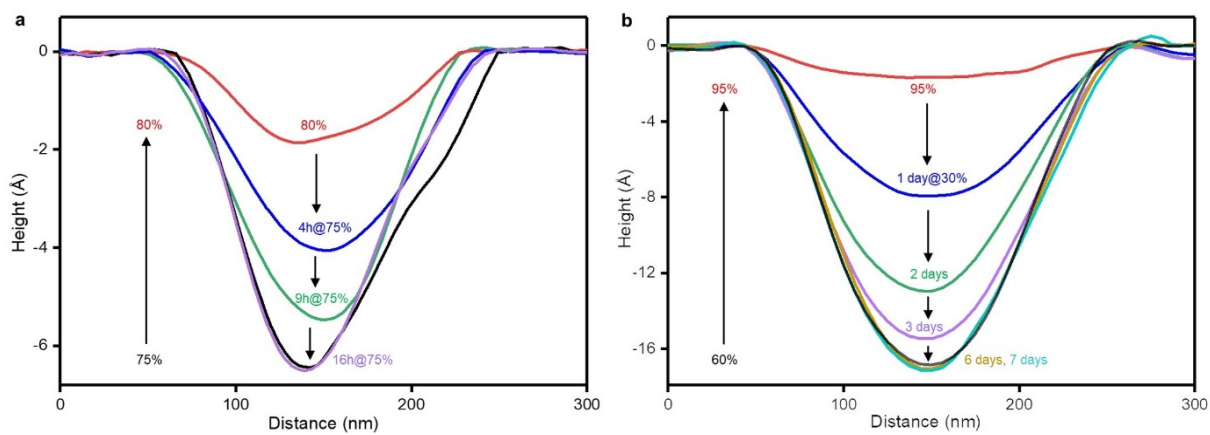

**Figure S4 | Non-hysteretic capillary condensation with slow dynamics. a,** Sagging profiles for a graphite capillary (N = 4) with increasing and decreasing RH between 75 and 80%. Black curve: initial dry-state profile. Red curve: RH was increased to 80%. Then, RH was returned to 75% and maintained at this humidity. AFM profiles were taken after 4, 9 and 16 h (color coded). **b,** The N = 6 graphite capillary was brought from the dry state (black curve) into the one filled with water and kept for an hour at 95% RH (red). The humidity was then decreased to ~30%, well below the condensation transition observed at 62.5 ±2.5% for this device. The color-coded curves show the time evolution towards the original dry state. Note that the sagging depths δ for such hysteretic loops were highly reproducible but details of sagging profiles could differ in different RH cycles. For example, the top crystal's adhesion to the right wall was different in the original and final dry states, as seen in **a** (cf. black and purple curves). This hysteresis is attributed to irreproducible vdW attachments of top crystals to channels' sidewalls.

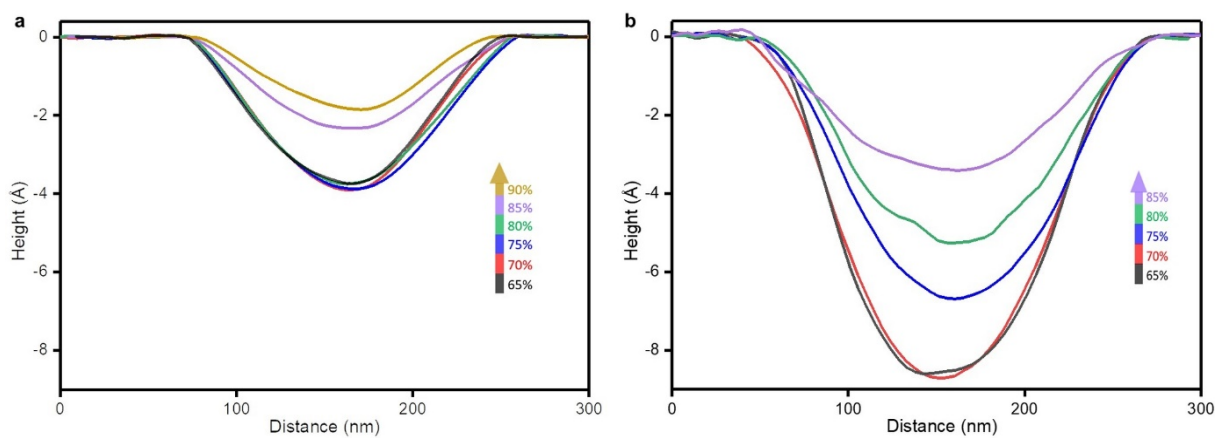

**Figure S5| Capillary condensation in 2D channels with different initial sagging.** Sagging profiles for two N = 5 graphite capillaries with different $\delta_0$. RH was increased in 5% steps (color coded). The water condensation transition occurred between 80 and 85% RH in **a** and between 70 and 75% in **b**. The difference in $RH_C$ for the same N is attributed to different h in the two cases.

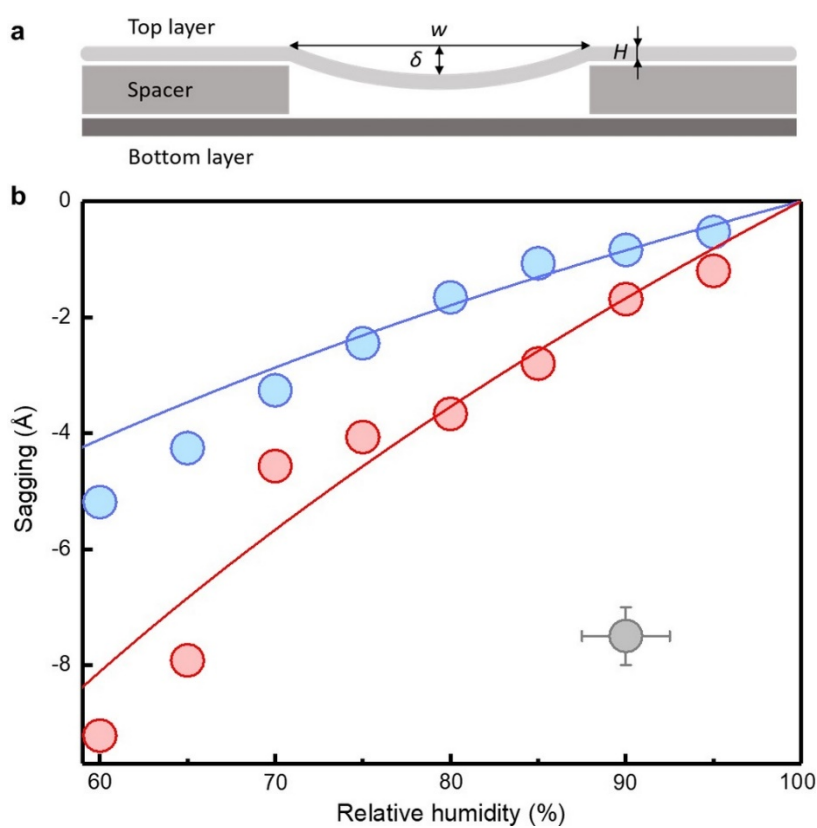

**Figure S6 | Remnant sagging above the condensation transition. a,** Schematic of top crystal's sagging. **b,** Typical behavior observed for the sagging depth $\delta$ as a function of RH, after the condensation transition occurred at RH < 60%. Symbols: Measurements for two different mica capillaries with N = 8. The solid curves are best fits using equations 3 and 4 (color-coded). The grey symbol with error bars indicates the experimental accuracy.

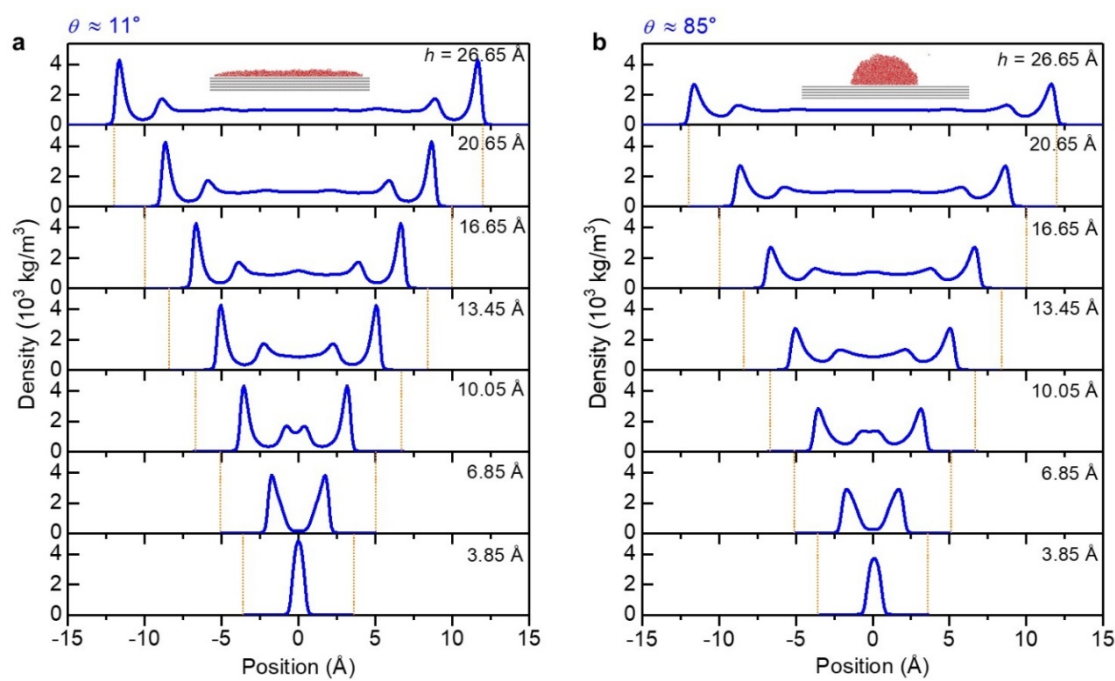

**Figure S7 | MD simulations of strongly confined water. a,** Its density profiles at different distances h between two rigid capillary walls with the contact angle $\theta \approx 11°$. **b,** Same calculations but for 85°. The orange dashed lines mark positions of the graphite surfaces that defined the 2D channels. Water exhibits a pronounced layered structure near each surface, and the structures start to overlap for h < 15 Å. Top insets: cross-sectional profiles for water droplets placed on the surfaces with the given $\theta$.

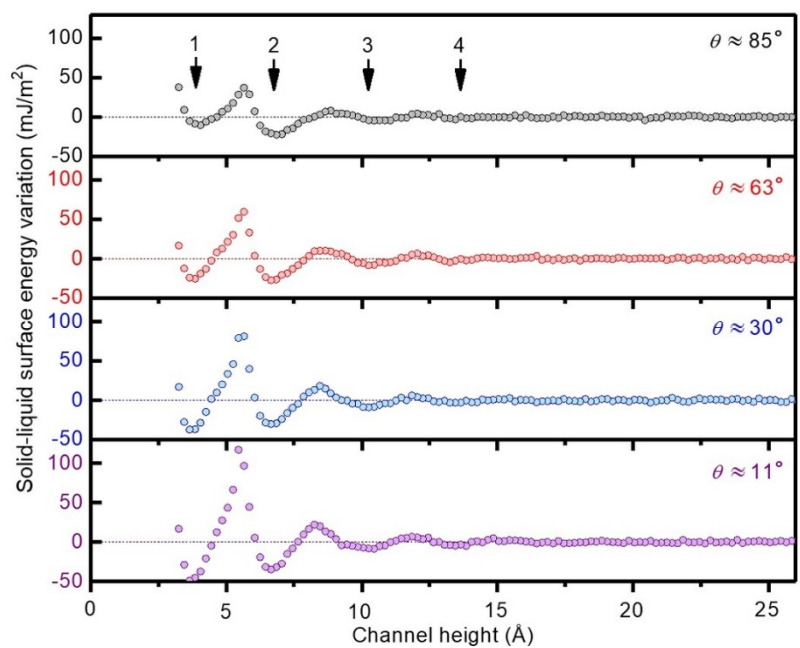

**Figure S8| Changes in the solid-liquid surface energy caused by atomic-scale confinement.** Calculated $\Delta\gamma$ (h) for several characteristic $\theta$. The arrows indicate the number of molecular layers of water, which fit inside the 2D channels.

16